\documentclass[prl,aps,twocolumn]{revtex4-1}
\usepackage{graphicx}
\usepackage{amsmath}
\usepackage{bm}
\usepackage{braket}
\usepackage{psfrag} 
\usepackage{latexsym} 
\usepackage{amstext}
\usepackage{amsxtra} 
\usepackage{float}
\usepackage{upgreek}
\usepackage{siunitx}
\usepackage{times}
\DeclareSIUnit\gauss{G}

\begin{document}
\newcommand{\To}{T_c^0}
\newcommand{\kB}{k_{\rm B}}
\newcommand{\dT}{\Delta T_c}
\newcommand{\lo}{\lambda_0}
\newcommand{\cs}{$\clubsuit \; $}
\newcommand{\thold}{t_{\rm hold}}
\newcommand{\Nmf}{N_c^{\rm MF}} 
\newcommand{\LM}{L_3^{\rm M}}
\newcommand{\Tmf}{T_c^{\rm MF}}
\newcommand{\downstate}{\left\vert\downarrow\right\rangle}  
\newcommand{\upstate}{\left\vert\uparrow\right\rangle}
\newcommand{\Ndown}{N_{\downarrow}}
\newcommand{\Nup}{N_{\uparrow}}
\newcommand{\overbar}[1]{\mkern 1.5mu\overline{\mkern-1.5mu#1\mkern-1.5mu}\mkern 1.5mu}

\newcommand{\kSymbol}{^{39}\mathrm{K}}
\newcommand{\intState}{\ket{1,1}}
\newcommand{\refState}{\ket{1,0}}
\newcommand{\dee}{\mathrm{d}}
\newcommand{\ee}{\mathrm{e}}

\newcommand{\omegaR}{\omega_r}
\newcommand{\omegaZ}{\omega_z}
\newcommand{\omegaI}{\omega_i}
\newcommand{\gammaElastic}{\gamma_{\textrm{el}}}
\newcommand{\gammaBoltzmann}{\gamma_{\textrm{r}}}

\newcommand{\upket}{\ket{\uparrow}}
\newcommand{\downket}{\ket{\downarrow}}
\newcommand{\uup}{\uparrow\uparrow}
\newcommand{\ddown}{\downarrow\downarrow}

\newcommand{\aB}{a_{\textrm{B}}}

\title{
Elliptic flow in a strongly-interacting normal Bose gas
}

\author{Richard~J.~Fletcher,$^{1,2}$$^\dagger$ Jay~Man,$^{1}$ Raphael~Lopes,$^{1}$ Panagiotis~Christodoulou,$^{1}$ Julian~Schmitt,$^{1}$ Maximilian~Sohmen,$^{1}$ Nir~Navon,$^{1,3}$ Robert~P.~Smith,$^{1}$ and Zoran~Hadzibabic$^{1}$}
\affiliation{$^1$ Cavendish Laboratory, University of Cambridge, J.~J.~Thomson Avenue, Cambridge CB3~0HE, United Kingdom\\
$^2$ MIT-Harvard Center for Ultracold Atoms, Research Laboratory of Electronics and Department
of Physics,\\Massachusetts Institute of Technology, Cambridge, Massachusetts 02139, USA\\
$^3$ Departments of Physics, Yale University, New Haven, Connecticut 06520, USA
}

\begin{abstract}
We study the anisotropic, elliptic expansion of a thermal atomic Bose gas released from an anisotropic trapping potential, for a wide range of interaction strengths across a Feshbach resonance.
We show that in our system this hydrodynamic phenomenon is for all interaction strengths fully described by a microscopic kinetic model with no free parameters. 
The success of this description crucially relies on taking into account the reduced thermalising power of elastic collisions in a strongly interacting gas, for which we derive an analytical theory.
We also perform time-resolved measurements that directly reveal the dynamics of the energy transfer between the different expansion axes.

\end{abstract}

\date{\today}


\maketitle

Elliptic flow, the collisional redistribution of energy between axes during expansion of a fluid, is a canonical example of hydrodynamic behaviour. Commonly discussed in the context of heavy-ion collisions~\cite{snellings:2011}, it has become a crucial probe of the strongly interacting quark-gluon plasma produced in these experiments. Due to the complicated microscopic physics, such systems are often described in terms of macroscopic (coarse-grained) quasi-equilibrium local quantities such as density and flow velocity. In this approach, the microscopic physics is encapsulated in phenomenologically introduced parameters such as viscosity.

Ultracold atomic gases offer an excellent testbed for studying the collective behaviour in interacting fluids, in large part because the low-temperature two-body interactions, characterised by the s-wave scattering-length $a$, can be tuned via magnetic Feshbach resonances~\cite{Chin:2010}. 
Most famously, atomic gases can show anisotropic expansion due to superfluid hydrodynamics~\cite{Zwierlein:2013,Pitaevskii:2016}. However, for sufficiently strong interactions they can also display pronounced elliptic flow in their normal state, above the critical temperature for superfluidity. 
This effect has been extensively studied in normal degenerate Fermi gases~\cite{OHara:2002a,Regal:2003a,Bourdel:2003,Trenkwalder:2011,Cao:2011,Elliott:2014a,Elliott:2014b,Joseph:2015,schaefer:2014}.
In normal Bose systems, elliptic flow has been observed for relatively weak interactions~\cite{Shvarchuck:2003,Gerbier:2004c} and in dipolar~\cite{Tang:2016} gases, 
but a systematic study as a function of the interaction strength has been lacking. Of particular interest is the hydrodynamic behaviour of the unitary Bose gas~\cite{leclair:2011,Chevy:2016,Klauss:2017,Eigen:2017}, in which $a \rightarrow \infty$ and the interactions are as strong as theoretically allowed.

\begin{figure}[t] 
	\includegraphics[width=1\columnwidth]{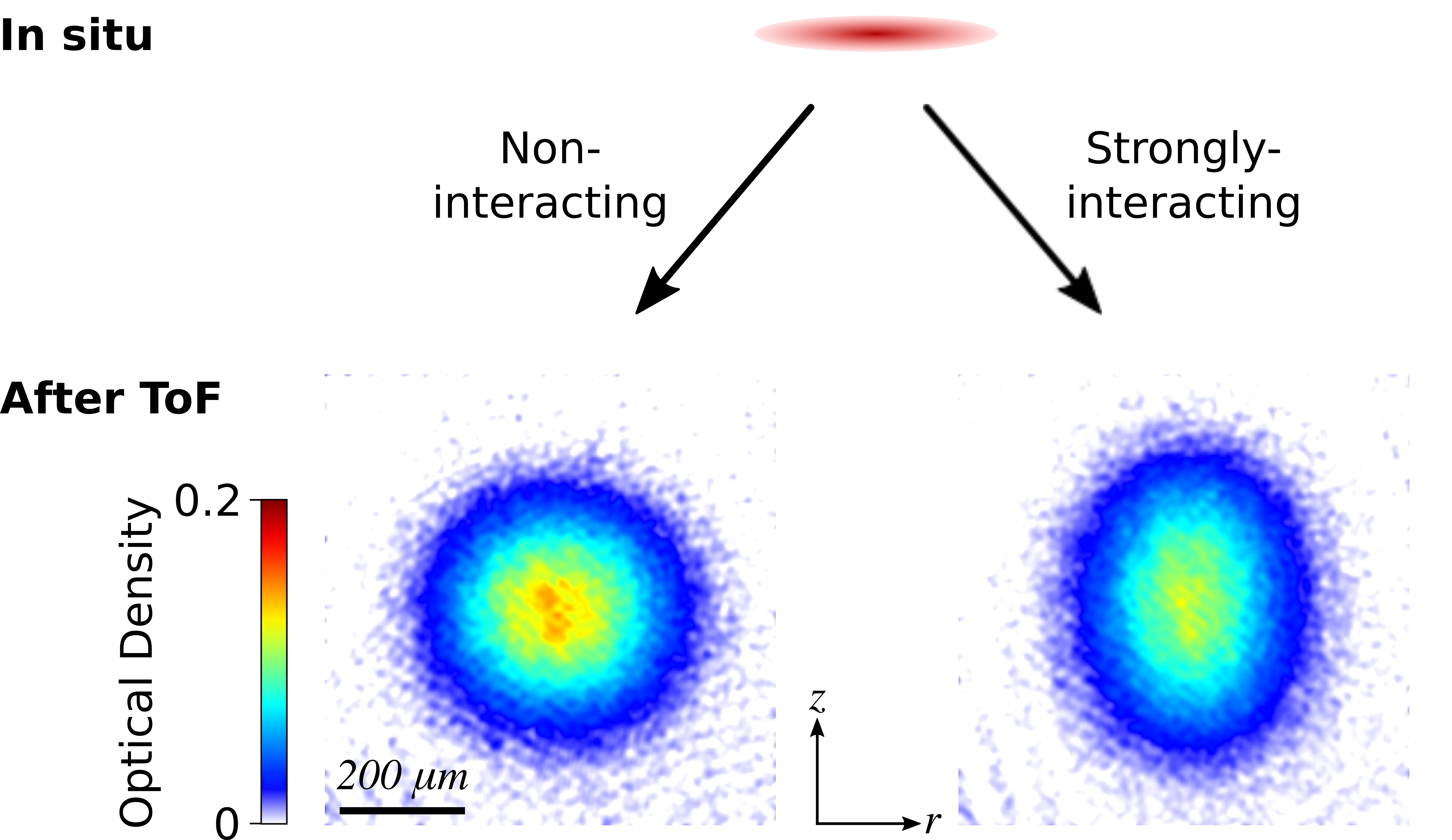}
	\caption{
Elliptic flow in a normal atomic Bose gas. Hydrodynamic behaviour is seen in the inversion of a cloud's spatial aspect ratio during time-of-flight (ToF) expansion, after release from an anisotropic harmonic trap. The sketch on top illustrates an anisotropic density profile of a trapped gas, whose parameters are given in the left panel of Fig.~\ref{fig:fig2}. The bottom panels show absorption images of the cloud after $10~$ms expansion, for two different interaction strengths. A quasi-ideal gas (left) expands essentially isotropically, while a strongly interacting one (right) shows a pronounced aspect-ratio inversion. Note that the trapped-gas cartoon is not to scale; for these experiments the aspect ratio of the trapped cloud was $\eta = 24$.}
	\label{fig:fig1}
\end{figure}

In this Letter, we study the elliptic flow of a normal atomic Bose gas, released from an anisotropic harmonic trap (see Fig.~\ref{fig:fig1}), for a wide range of interaction strengths across a Feshbach resonance, and a wide range of trap anisotropies. 
We show that despite being a quintessentially hydrodynamic phenomenon, elliptic flow in our system can in all interaction regimes be described
by a microscopic kinetic model with no phenomenological parameters.  
To explain our observations for $a\rightarrow \infty$, it is crucial to take into account not only the unitarity-imposed limitations on the scattering rate, 
but also on the effectiveness of collisions in transferring energy between the expansion axes, 
for which we derive an analytical theory. 
Exploiting the possibility to turn the interactions on and off at any point during the gas expansion, we also perform time-resolved experiments that directly reveal the dynamics of the energy transfer between the expansion axes.

\begin{figure*}[tb] 
	\includegraphics[width=\textwidth]{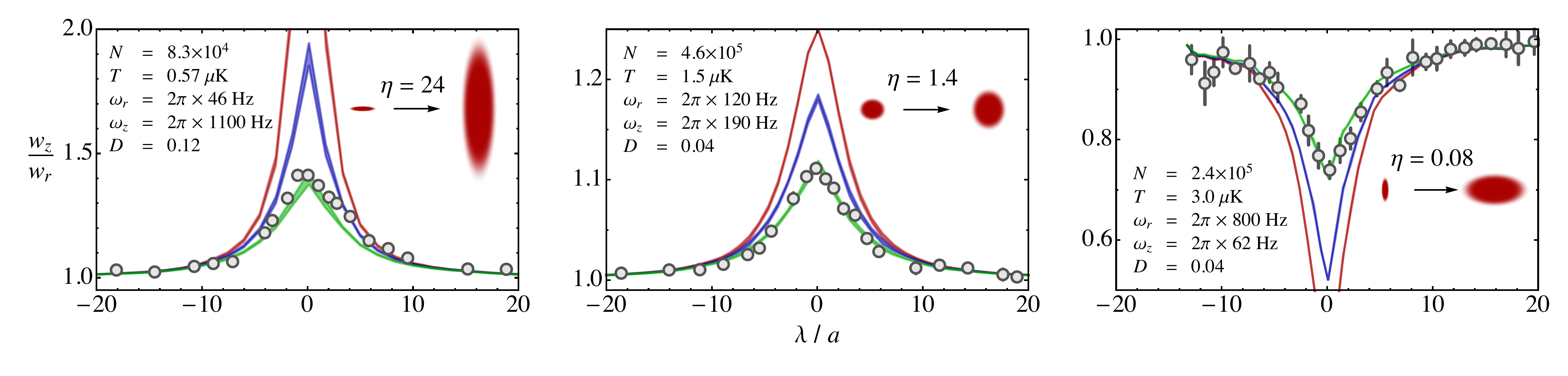}
	\caption{
		Spatial aspect ratio of the cloud after expansion, $w_z/w_r$, for various interaction strengths across a Feshbach resonance, and different trap geometries. For each data set (panel) the inset cartoons (not to scale) indicate the trap geometry characterised by $\eta = \omega_z/\omega_r$, the ratio of axial and radial trapping frequencies in our cylindrically symmetric harmonic trap. The legends also show the trapped-cloud atom number $N$, temperature $T$, and peak phase-space density $\mathcal{D}$. In each panel we show three different theoretical curves, corresponding to progressively more refined models. In red we show the theory for a weakly interacting gas, in blue a theory that accounts for the unitarity-imposed limitations on the elastic scattering rate, and finally in green our theory that also takes into account the reduced ability of collisions to transfer momentum in a strongly interacting Bose gas (see the text for details). The shading of the theoretical curves reflects the variations and uncertainties in $N$ and $T$, and also the atom-number uncertainty due to small  ($<8\%$) three-body losses during the expansion.}
	\label{fig:fig2}
\end{figure*}

Our experimental setup is described in \cite{Campbell:2010}. We work with $\kSymbol$ atoms trapped in a cylindrically-symmetric optical harmonic potential, with a tuneable trap anisotropy $\eta = \omegaZ/\omegaR$, where $\omega_{z,r}$ are the axial and radial trapping frequencies. We use two internal (spin) states, $\upket = \ket{F=1,m_F=1}$ and $\downket = \ket{F=1,m_F=0}$, labelled in the low-field basis, and tune the $\uup$ scattering length $a$ using the magnetic Feshbach resonance centred at $\SI{402.70(3)}{\gauss}$~\cite{Fletcher:2017}. The $\ddown$ scattering length is negligible ($<10~a_0$, where $a_0$ is the Bohr radius) for all relevant magnetic field strengths. 
The peak phase-space density in our trapped clouds, $\mathcal{D} =  n_0  \lambda^3$, is $\lesssim 0.1$ (see Fig.~\ref{fig:fig2} legends), 
so they are well described by gaussian real- and momentum-space distributions; here $n_0$ is the number density in the centre of the cloud and $\lambda=h/\sqrt{2\pi m\kB T}$ is the thermal wavelength, with $T$ the temperature and $m$ the particle mass. 

\begin{figure}[t]
	\centering
		\includegraphics[width=1\columnwidth]{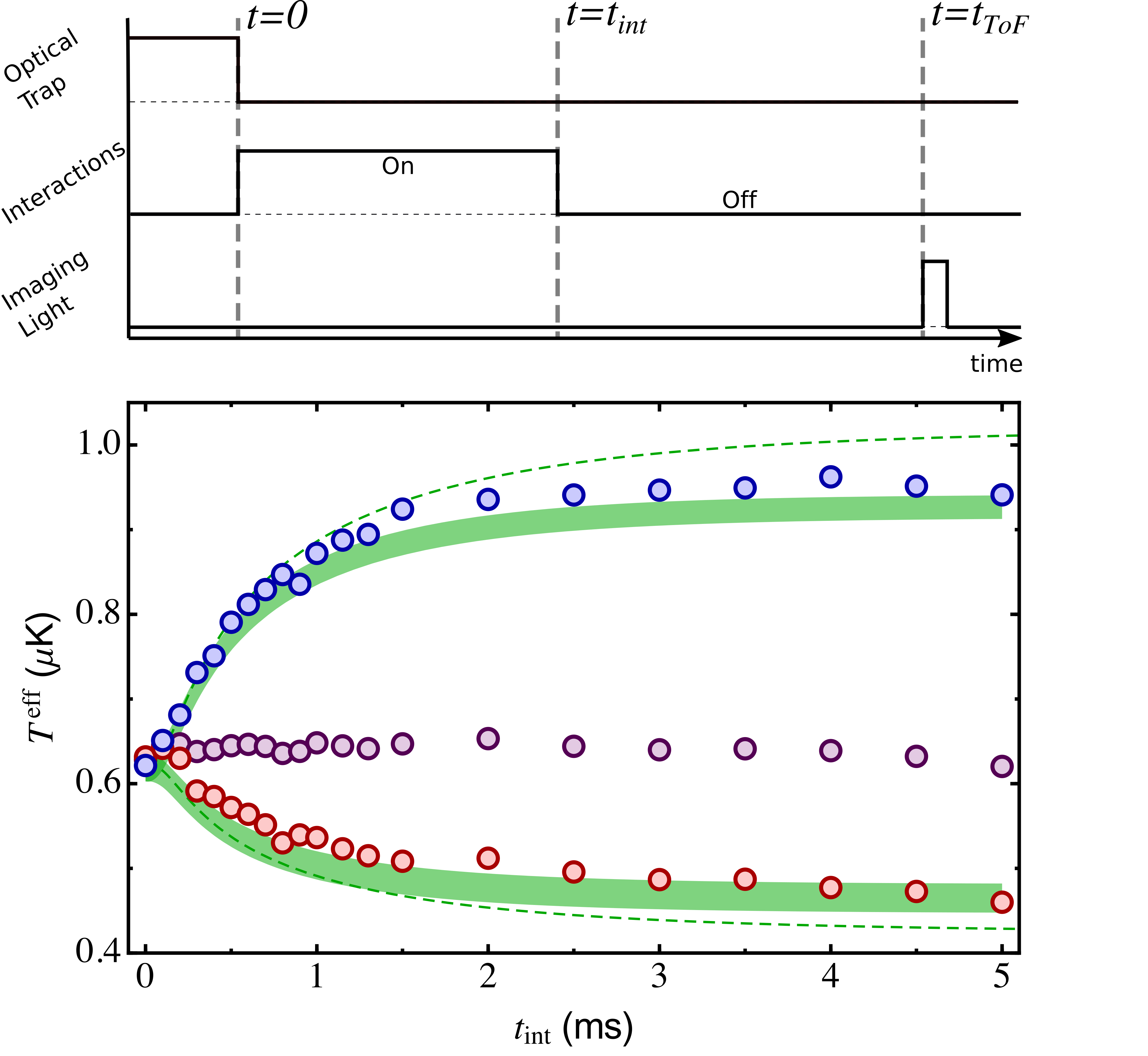}
	\caption{Dynamics of the inter-axis energy transfer in the early stages of the expansion. The top panel illustrates our experimental protocol for obtaining time-resolved measurements of the momentum distribution. In the bottom panel we show the effective temperature $T^\textrm{eff}$ associated with motion along the axial (blue) and radial (red) directions during expansion of a resonantly-interacting gas ($\lambda/a =0$) released from a strongly oblate trap ($\eta = 24$); all the trapped-cloud parameters are essentially the same as in the left panel of Fig.~\ref{fig:fig2}. In purple we show the average temperature, $(2T^\textrm{eff}_r+T^\textrm{eff}_z)/3$, which remains essentially constant. The green bands show numerical simulations for our experimental parameters, with the band thickness reflecting the atom-number and temperature uncertainties. The dashed lines show the predictions for an ideal experiment with infinite time of flight.}
\label{fig:fig3}
\end{figure}

We prepare a quasi-ideal equilibrium gas in $\downket$~\cite{Fletcher:2017}, then release the cloud from the trap and simultaneously transfer it to $\upket$ with a radio-frequency (RF) $\pi-$pulse~\footnote{The Rabi frequency, $\Omega\approx 2\pi\times 15~$kHz, of our RF drive is much larger than the interaction energy of the final cloud, resulting in a transfer efficiency of essentially unity.}. This $34$-$\mu$s pulse is very short compared to our characteristic millisecond expansion timescale, set by $\omega_\textrm{max} = \max[\omega_r, \omega_z]$ (see Fig.~\ref{fig:fig2} legends). Hence, our RF spin flip acts as an essentially instantaneous interaction switch, and from the start of the expansion the (local) rate of elastic collisions is $\gamma_\textrm{el} = n \langle \sigma \hbar k/m \rangle$, where $n$ is the density, $\sigma=8\pi a^2/(1+k^2 a^2)$ is the (unitarity-limited) scattering cross section, $\hbar k$ is the relative momentum of the particles, and $\langle\dots\rangle$ denotes a thermal average. Finally, after $\SI{10}{\milli\second}$ of ToF we image the cloud radially (see Fig.~\ref{fig:fig1}), and extract its axial and radial gaussian widths, $w_{z,r}$
~\footnote{For some of the measurements the interactions are turned off by switching off the Feshbach field after $5$~ms of ToF, by which time the collision rate is already negligible; see Fig.~\ref{fig:fig3} later.}.
We normalise the measured aspect ratio to that obtained by repeating the experiment with the quasi-ideal $\downket$ gas ({\it i.e.}, omitting the RF spin-flip pulse), which removes small (few percent) systematic anisotropies due to imaging artefacts and the non-infinite ToF.

In Fig.~\ref{fig:fig2} we show the aspect ratio $w_z/w_r$ measured for various scattering lengths across the Feshbach resonance, and (in different panels) for widely different trap geometries, from strongly oblate ($\eta \approx 24$) to strongly prolate ($\eta \approx 0.08$). 
For all our values of $\eta$, we observe the expected quasi-isotropic expansion for weak interactions, and the cloud expands most anisotropically at unitarity, where the elastic scattering rate is maximal. 

Hydrodynamic behaviour should be pronounced if the initial scattering rate in the cloud centre, $\gamma_\textrm{el}^0=n_0 \langle \sigma \hbar k/m \rangle$, is much larger than the expansion rate $\omega_\textrm{max}$~\footnote{Note that the ratio $\gamma_\textrm{el}^0/\omega_\textrm{max}$ is essentially the inverse of the Knudsen number, defined as the ratio of the mean free path to the system size~\cite{Shvarchuck:2003}}. In our unitary clouds, $\gamma_\textrm{el}^0/\omega_\textrm{max}$ 
varies between $5$ (for the $\eta=24$ data) and as much as $24$ (for the $\eta=1.5$ data),
consistent with the observed pronounced elliptic flow. 
At the same time, since $\mathcal{D} \ll 1$, even at unitarity the mean free path  $\ell \sim (n_0\lambda^2)^{-1} \sim \lambda/\mathcal{D}$ is much larger than the de Broglie wavelength of the particles ($\lambda$) and the typical interparticle distance $n_0^{-1/3} \sim \ell \mathcal{D}^{2/3}$. This means that particles still have a well-defined momentum and we can primarily consider their pairwise interactions. 
We thus model our experiments using the Boltzmann equation for the semi-classical phase-space distribution $f(\boldsymbol{r},\boldsymbol{v},t)$, which gives the occupation of the phase-space element corresponding to spatial position $\boldsymbol{r}$ and velocity $\boldsymbol{v}$:
\begin{equation}
\frac{\partial f}{\partial t}+ \boldsymbol{v} \cdot\frac{\partial f}{\partial \boldsymbol{r}}=-\gammaBoltzmann (f-f_\textrm{le}).
\label{eqn:boltzmann}
\end{equation}
The left-hand side of Eq.~(\ref{eqn:boltzmann}) is the convective derivative of $f$, while the right-hand side is the Boltzmann collision integral in the relaxation-time approximation, which assumes that $f$ relaxes to its local equilibrium value $f_\textrm{le}$ at a rate $\gammaBoltzmann$; the (instantaneous) $f_\textrm{le}$ corresponds to an isotropic, thermal momentum distribution in the zero-momentum frame for all the particles at a given $\textbf{r}$, with the same local kinetic energy density as for $f$~\cite{Pedri:2003}.
For a cloud released from an anisotropic trap, this {\it local} thermalisation results in a transfer of energy from the low-$\omega$ direction(s) to the high-$\omega$ one(s)~\cite{Pedri:2003}.
By adopting a gaussian ansatz for $f$, one can obtain and solve equations of motion for its real- and momentum-space widths, which fully define the phase-space distribution; in this approach $\gammaBoltzmann$ becomes a global variable that can be related to the elastic collision rate averaged across the cloud, $\overline{\gammaElastic}$~\cite{Pedri:2003,Guery-Odelin:1999a}.

We compare our data to three different theoretical curves, indicated by different colours in Fig.~\ref{fig:fig2}, which correspond to progressively better models for $\gammaBoltzmann$. The red curve shows the weakly-interacting theory~\cite{Pedri:2003}, where one approximates $\sigma = 8\pi a^2$, so $\overline{\gammaElastic}=(8\pi a^2\hbar/m)\left<n k\right>$ and for this case it was shown that $\gammaBoltzmann=(4/5)\overline{\gammaElastic}$~\cite{Guery-Odelin:1999a}. 
This approach works well for relatively weak interactions, but as expected fails at and near unitarity, since it unphysically assumes $\sigma \rightarrow \infty$ for $a\rightarrow \infty$. 
The blue-curve calculation properly takes into account the unitary saturation of the scattering cross-section, so $\overline{\gammaElastic}=(8\pi\hbar a^2/m)\left<n k/(1+k^2 a^2)\right>$, but we still assume $\gammaBoltzmann=(4/5)\overline{\gammaElastic}$ and still overestimate the cloud anisotropy at and near unitarity. 
The reason for this is that the ratio $\gammaBoltzmann/\overline{\gammaElastic}$ is also affected by the $k$-dependence of $\sigma$. At unitarity, a larger fraction of collisions contributing to the total $\overline{\gammaElastic}$ occurs between particles with small relative momenta; such collisions are not effective in thermalising the gas, and consequently $\gammaBoltzmann/\overline{\gammaElastic}$ is reduced. This effect was previously discussed in the context of in-trap cross-dimensional thermalisation, and it was numerically found that the thermalisation rate at fixed $\gammaElastic$ is suppressed by a factor of up to 4~\cite{Arndt:1997,Kavoulakis:2000c}. Here, we generalise the calculation of~\cite{Guery-Odelin:1999a} to include the $k$-dependence of $\sigma$ and derive an analytic expression for $\gammaBoltzmann/\overline{\gammaElastic}$ valid at all interaction strengths. Writing $\gammaBoltzmann/\overline{\gammaElastic} = (4/5) g^{-1}(\alpha)$, where
$\alpha = 2 \pi a^2/\lambda^2$, we get
\begin{equation}
g(\alpha) = 6 \int_0^{\infty} \frac{ x^3 \, e^{-x^2}}{1 + \alpha x^2} \, {\rm d}x \, \left[ \int_0^{\infty} \frac{ x^7 \, e^{-x^2}}{1 + \alpha x^2} \, {\rm d}x \right]^{-1} \, .
\end{equation}
For $\alpha = 0$ we recover $\gammaBoltzmann/\overline{\gammaElastic} = 4/5$, while for $\alpha \rightarrow \infty$ we get $g=3$ and $\gammaBoltzmann/\overline{\gammaElastic} = 4/15$, meaning that relaxation requires three times as many collisions.
The resulting prediction for the aspect ratio of the expanding cloud is shown in green in Fig.~\ref{fig:fig2}. Without any free parameters, we capture the experimental data excellently for all interaction strengths and for values of $\eta$ differing by more than two orders of magnitude.

Finally, in the last part of the paper, we use our spin-flip interaction switch to experimentally  time-resolve the transfer of energy between the different expansion axes and more directly reveal the underlying mechanism for the elliptic flow (see Fig.~\ref{fig:fig3}). Here we focus on a unitary Bose gas and a strongly oblate trap with $\eta = 24$.
As before, at time $t=0$ we release the cloud from the trap and turn on the interactions by spin-flipping it from $\downket$ to $\upket$. 
However, now, after some variable short interaction time, $t_{\textrm{int}} < t_\textrm{ToF}=10~$ms, we turn off the interactions by spin-flipping the cloud back to $\downket$ and thus suddenly interrupt the energy transfer. For the remainder of ToF the expansion proceeds ballistically and the final shape of the cloud reflects the momentum distribution frozen at $t= t_{\textrm{int}}$.

The $t_{\textrm{int}}$-dependent expansion energy along each direction is characterised by an effective temperature, $T_{i}^\textrm{eff}$, where $i = r, z$; this {\it global} $T_{i}^\textrm{eff}$ corresponds to projecting the phase-space distribution onto the $i$ momentum axis.
 In Fig.~\ref{fig:fig3}, we show $T_{i}^\textrm{eff} (t_{\textrm{int}})$ that we approximately extract from the measured cloud widths after ToF using
\begin{equation}
T^\textrm{eff}_{i} \approx \frac{m}{\kB}\frac{w_{i}^2-w_{0,i}^2}{t_\textrm{ToF}^2}  \, ,
\label{eqn:effT}
\end{equation}
where $w_{0,i}=\sqrt{\kB T/(m\omega_{i}^2)}$ are the spatial sizes of the trapped cloud; note that the relationship in Eq.~(\ref{eqn:effT}) would be an exact equality for $t_\textrm{ToF} \rightarrow \infty$.
At the time of release $T^\textrm{eff}_{r}=T^\textrm{eff}_{z}$, but as the expansion proceeds in the presence of interactions, $T^\textrm{eff}_{r}$ (red) decreases and $T^\textrm{eff}_{z}$ (blue) increases. The mean temperature, $(2T^\textrm{eff}_{r}+T^\textrm{eff}_{z})/3$ (purple), remains constant, confirming an elastic redistribution of energy between the different axes. 

We note that the energy redistribution process extends over a characteristic timescale of $\approx 1$~ms, which is notably longer than the ideal-gas expansion timescale $1/\omega_z=0.14~$ms. One reason for this is that at unitarity the drop in $\gamma_\textrm{el}$ due to the drop in the density during expansion is countered by an increase due to the reduction in the local spread of momenta, which increases the unitarity-limited $\sigma$ (for a related discussion see also~\cite{Bourdel:2003}).

Here we again compare the experimental data with our numerical simulations (with $\gammaBoltzmann/\overline{\gammaElastic} = 4/15$) and find good agreement without any free parameters. The green bands in Fig.~\ref{fig:fig3} show simulations for our experimental protocol and parameters; that is, we directly simulate the right hand side of Eq.~(\ref{eqn:effT}) rather than the exact $T_{i}^\textrm{eff}$. For comparison, with dashed lines we also show simulations of the exact $T_{i}^\textrm{eff}$, which would be observed in an ideal experiment with $t_\textrm{ToF} \rightarrow \infty$, and find that our measurements are close to these predictions.

In conclusion, we have studied elliptic flow in a normal Bose gas with tuneable interactions, and have quantitatively explained this quintessentially  hydrodynamic behaviour using a microscopic kinetic model with no free parameters.
Our measurements show that the behaviour of a strongly interacting gas is crucially affected by the reduction of the effectiveness of elastic collisions in driving the system towards local equilibrium, for which we have derived an analytical theory.
Finally, by studying the time-dependence of the expanding cloud's momentum distribution, we have directly revealed the dynamics of the energy transfer between the expansion axes.

We thank John Thomas for useful discussions.
This work was supported by EPSRC [Grants No. EP/N011759/1 and No. EP/P009565/1], ERC (QBox), AFOSR, and ARO. R.J.F. acknowledges support from the Pappalardo Fellowship. R.L. acknowledges support from the E.U. Marie-Curie program [Grant No. MSCA-IF-2015 704832] and Churchill College, Cambridge. N.N. acknowledges support from Trinity College, Cambridge. R.P.S. acknowledges support from the Royal Society. 

$^\dagger$ rfletch@mit.edu



\end{document}